%% file: fermions.tex
\documentclass[12pt]{iopart_mod}

\usepackage{amsfonts,amssymb,amsmath,amsthm,epsfig,cite}
\usepackage{array}

\input{macros}

\begin{document}

\hfill {\footnotesize LU-ITP 2006/010}

\title{Quantum Dilaton Gravity in Two Dimensions with Fermionic Matter}

\author{D.~Grumiller and R.~Meyer}

\address{Institut f\"ur Theoretische Physik, Universit\"at Leipzig, Augustusplatz 10-11, D-04109 Leipzig, Germany} 

\eads{\mailto{grumil@lns.mit.edu}, \mailto{Rene.Meyer@itp.uni-leipzig.de}}

\begin{abstract}
  Path integral quantization of generic two-dimensional dilaton
  gravity non-minimally coupled to a Dirac fermion is performed. After
  integrating out geometry exactly, perturbation theory is employed in
  the matter sector to derive the lowest order gravitational vertices.
  Consistency with the case of scalar matter is found and issues of
  relevance for bosonisation are pointed out.
\end{abstract}


\setcounter{footnote}{0}

\section{Introduction}

Two-dimensional gravity models 
retain many properties of their higher-di\-men\-sio\-nal
counterparts but are considerably simpler and thus can be used as
toy models for tackling the conceptual problems of quantum gravity
while at the same time avoiding the technical difficulties that arise
from the non-linear dynamics of gravity theories in higher
dimensions. The most prominent of them, the Schwarzschild black hole, is not merely a toy model but of considerable interest for General Relativity in four (or higher) dimensions. 
Two-dimensional dilaton gravities
\cite{Grumiller:2002nm} (cf.~also \cite{Brown:1988} 
for earlier reviews), being not only
classically integrable but even at
the (non-perturbative) quantum level, cf.~e.g.~\cite{Kummer:1996hy,Kummer:1997jj,Kummer:1998zs,Grumiller:2001ea,Bergamin:2004us,Cangemi:1995yz}, 
especially fit
that purpose, such that it is possible to discuss topics like
background independence \cite{Grumiller:2003sk}, the role of time
\cite{Schaller:1992np} and the scattering on virtual black holes \cite{Grumiller:2004yq}.

In this work we investigate the first order formulation
\cite{Schaller:1994es} 
of two-dimension\-al dilaton gravities coupled to fermions in the
``Vienna School approach''
\cite{Grumiller:2002nm,Kummer:1992bg}. 
Such models were first considered in the second order formalism
classically in \cite{Cavaglia:1998yp} and used later on in studies
\cite{Nojiri:1992st,Ori:2001xc} of the evaporation of charged CGHS
black holes \cite{Callan:1992rs}.  Interesting by itself because they
have been a blind spot in the literature on quantum dilaton gravity
until now, the main motivation of our work is to provide the grounds
for an investigation of bosonisation, i.e.~the quantum equivalence
between the massive Thirring model and the Sine-Gordon model
\cite{Coleman:1975pw}, 
in the context of quantum dilaton gravity. Originally, this
equivalence has been stated for the corresponding field theories on
two-dimensional Minkowski space and used in recent studies
\cite{Frolov:2005ps,Frolov:2006is,Thorlacius:2006tf} of charged black hole evaporation
in two-dimensional dilaton gravity electrodynamics, i.e.~on a fixed
background with a quantized matter sector, where it is applicable in
regions of small curvature (compared to the intrinsic length scale of
the quantum theory). As a non-perturbative quantization of
two-dimensional dilaton gravity theories coupled to scalar matter is
already available \cite{Kummer:1997jj,Kummer:1998zs,Grumiller:2001ea},
the question arises whether and how bosonisation carries over to the
quantum gravity regime. A necessary prerequisite for answering this
question is to perform the same constraint analysis
\cite{Meyer:2005fz} and -- the main topic of this work -- exact path
integral quantization analogously to the scalar case. In a next step
one can then compare the physical observables (e.g.~S-matrix elements)
on both sides of the correspondence. The underlying rationale is to
integrate out geometry without split into background and fluctuations.
The ensuing effective theory will be non-local and non-polynomial in
matter degrees of freedom and can be studied with standard
perturbation theory. To each order all gravitational backreactions are
included automatically in a self-consistent way.

This paper is organised as follows: In section \ref{se:2} the starting
point, dilaton gravity in two dimensions with fermionic matter, is
provided and our notation is introduced. Section \ref{se:3} is devoted
to a Hamiltonian analysis of constraints, the construction of the BRST
charge and the gauge fixing procedure. The path integral quantization
of geometry is performed non-perturbatively in section \ref{se:4}.
Some applications of the general results are given in section
\ref{se:5}, in particular the derivation of four-point vertices. We
conclude in section \ref{se:6} with a study of conformal properties,
comments on bosonisation and an outlook to further possible
applications.

\section{Two-Dimensional Gravity with Fermions}\label{se:2}

Our starting point is the two-dimensional (2D) action
\begin{equation}
S=S^{(1)}+S^{(\rm kin)}+S^{(\rm SI)} 
\label{eq:start}
\end{equation}
which comprises the first order action of 2D dilaton gravity,
\beqa\label{eq:FOG}
S^{\rm (1)} & = & \int_{\mathcal{M}_2} \left[X_aT^a+X\cR+\epsilon\mathcal{V} (X^aX_a,X)\right] \,, 
\eeqa
the Dirac action\footnote{The sign difference between
  \cite{Meyer:2005fz} and this article for $S^{(\rm kin)}$ stems from the
  differing sign of $\eps^{ab}$.}
\beqa\label{eq:fermkin}
S^{(\rm kin)} & = & \frac{i}{2} \int_{{\cal M}_2} f(X)\;(*e^a) \wedge (\chib \ga_a \overleftrightarrow{\extd}\chi)\,,
\eeqa
and fermion self-interactions
\beqa
S^{(\rm SI)}  & = & - \int_{{\cal M}_2} \epsilon h(X) g(\chib\chi)\label{eq:fermSI}\,.
\eeqa
The reader familiar with the notation used in \eqref{eq:FOG}-\eqref{eq:fermSI} may skip the rest of this section which is devoted to detailed explanations of the latter and also provides a brief recollection of some well-known results in 2D dilaton gravity. For background information and additional references the extensive
review \cite{Grumiller:2002nm} may be consulted.

\subsection{Notations and conventions}\label{se:2.1new}

We collect first all notations relevant for the geometric action \eqref{eq:FOG}. To formulate this first order action one needs to introduce
{\it Cartan} variables:
$e^a=e^a_\mu dx^\mu$ is the dyad one-form dual to $e_a = e_a^\mu \partial_\mu$,
i.e.~$e^a(e_b)=\de^a_b$.  Latin indices refer to an anholonomic frame,
Greek indices to a holonomic one. 
The Levi-Civita tensor is given by $\epsilon_{\mu\nu} = |\det e_\al^a| \tilde{\eps}_{\mu\nu}$ with $\tilde{\eps}_{01}=+1$. 
For calculations it is often
convenient to express everything in light-cone gauge for the flat metric $\eta_{ab}$,
\eq{
 \eta_{+-}=1=\eta_{-+}\,,\qquad \eta_{++}=0=\eta_{--}\,.
}{eq:eta}
The volume 2-form may be presented as 
\eq{
\epsilon =  - \frac{1}{2} \eps_{ab} e^a \wedge e^b = e^+ \wedge e^-=(e)\extd^2 x\,,\qquad (e):=\det e_\mu^a=e_0^+e_1^- - e_0^-e_1^+\,,
}{volform2}
with the totally
antisymmetric Levi-Civita symbol in tangent space $\eps_{ab}$ defined with the same sign as $\tilde{\eps}_{\mu\nu}$. Consequently, $\eps^a{}_b$ is simply given by $\eps^\pm{}_\pm=\pm 1$. 
The Hodge star acts on the dyad as
$\ast e^a = -\eps^a{}_b e^b$ and on the volume 2-form as $\ast\epsilon=1$.
With these conventions the hermitian
conjugate of the exterior derivative \cite{nakaharageometry} reads
$\extd^\dagger = \ast \extd \ast$.

The one-form $\omega$ represents the
spin connection $\om^a{}_b=\eps^a{}_b\om$. 
The torsion two-form in light-cone gauge for the anholonomic frame is given by
\eq{
T^\pm=(\extd\pm\omega)\wedge e^\pm\,.
}{eq:torsion}  
The curvature two-form
$\cR^a{}_b$ can be represented by the two-form $\cR$ defined by
$\cR^a{}_b=\eps^a{}_b \cR$, 
\eq{
\cR=\extd\om\,. 
}{eq:curvature}
The {\em Cartan} variables, $e^\pm$ and $
\om$, are the gauge field 1-forms entering the action \eqref{eq:FOG}, together with their ``field strengths'' \eqref{eq:torsion} and \eqref{eq:curvature}.

The fields $X, X^a$ (or in light-cone gauge $X,X^\pm$) are zero forms and may be interpreted as Lagrange multipliers for curvature and torsion, respectively. The quantity $\mathcal{V} (X^aX_a,X)$ is
 an arbitrary function of Lorentz invariant combinations of these Lagrange multipliers. Actually,
for most practical purposes the potential takes the simpler
form\footnote{To the best of our knowledge the only exception
  appearing in the literature is the class of dilaton-shift invariant
  models introduced in \cite{Grumiller:2002md} with ${\cal
    V}(X^+X^-,X)=X U(X^+X^-/X^2)$.}
\begin{equation}
  \label{eq:pot}
  \mathcal{V} (X^aX_a,X) = X^+X^- U(X) + V(X)\,.
\end{equation}
The functions $U,V$ are the crucial input defining the geometric part of the model, and several examples are presented below. The scalar field $X$ is called ``dilaton field'' for reasons pointed out in the next subsection.

We present now the missing pieces of notation required to comprehend the matter actions \eqref{eq:fermkin} and \eqref{eq:fermSI}. Let us start with the Dirac matrices in 2D Minkowski space ($\ga^\pm=(\ga^0\pm\ga^1)/\sqrt{2}$)
\eq{
\begin{array}{ll}
\ga^0 = \ \;\left(
\begin{array}{rr}
	0 & 1 \\
	1 & 0
\end{array}\right)
& \ \ \ 
\ga^1 = \left(
\begin{array}{rr}
	0 & 1 \\
	-1 & 0
\end{array}\right) \\
\ga^+ = \left(
\begin{array}{rr}
	0 & \sqrt{2} \\
	0 & 0
\end{array}\right)
& \ \ \ 
\ga^- = \left(
\begin{array}{rr}
	0 & 0 \\
	\sqrt{2} & 0
\end{array}\right)\,.
\end{array}
}{eq:diracmatrices}
The analogue of the $\ga^5$ matrix is $\ga_\ast = \ga_0 \ga_1 = {\rm
  diag}(+-)$. They satisfy $\{\ga^a,\ga^b\} = 2 \eta^{ab}$ and
$\{\ga_\ast,\ga^a\} = 0$. The Dirac conjugate is defined in the usual way, $\bar{\chi}=\chi^\dagger \ga^0$.
For calculations in Euclidean space $\ga^0$
is defined as above, but $\ga^1 = {\rm diag}(+-)$ and $\ga_\ast =
\ga_0\ga_1$, thus satisfying $\{\ga^a,\ga^b\} = 2 \de^{ab}$. The Dirac
matrices in Euclidean space are hermitian, $\ga^a =
{\ga^a}^\dagger$, whereas $\ga_\ast$ becomes anti-hermitian. 
The derivative action on both sides in \eqref{eq:fermkin}
  is defined as $a\lrpd{}b = a\partial b - (\partial a) b$. 

The functions $f(X)$ and $h(X)$ entail the coupling to the dilaton. If
they are constant the fermions are called minimally coupled, and
non-minimally coupled otherwise. Because of the Grassmann property of
the spinor field the self-interaction $g(\chib\chi)$ may be Taylor expanded as
\eq{
g(\chib\chi) = c + m \chib\chi +\lambda (\chib\chi)^2\,.
}{eq:fersi}
The constant contribution $c$ can be absorbed into $V(X)$. Thus, there
is only a mass term (if $m\neq 0$) and a Thirring term (if $\lambda\neq 0$). 
Also, because 
there is only one generator of the Lorentz group, the fermion kinetic
term does not include coupling to the spin connection, which together
with the requirement that matter does not couple to the Lagrange
multipliers for torsion $X^a$ is crucial for the equivalence between the
first and second order theories with matter
\cite{Katanaev:1995bh} 
and, as will be seen below, simplifies the constraint structure
significantly. 

The action \eqref{eq:start} depends functionally on the fields $X,X^\pm,\om,e^\pm$ and $\chi$. Due to the presence of gauge symmetries the only propagating physical degree of freedom in our
model is the two-component spinor $\chi$. It should be noted that although in general the addition of fermions (as well as other matter fields) destroys classical integrability of first order gravity, some special cases still can be treated exactly, for instance chiral fermions \cite{Kummer:1992ef}.

\subsection{Some properties and examples of 2D dilaton gravity}\label{se:2.2new}
The action \eqref{eq:FOG} is equivalent to the frequently used second
order action \cite{Russo:1992yg} 
\begin{equation}
\label{eq:GDT}
S^{(2)}=-\frac12 \int_{\mathcal{M}_2} \extd^{2}x\, \sqrt{-g}\; \left[ X R + U(X)\; (\nabla X)^{2} - 2V(X)\; \right] \, ,
\end{equation}
with the same functions $U,V$ as in \eqref{eq:pot}. The curvature
scalar\footnote{The sign of the curvature scalar $R$ has been fixed
  conveniently such that $R>0$ for $dS_2$. This is the only difference
  to the notations used in ref.~\cite{Grumiller:2002nm}.} $R$ and
covariant derivative $\nabla$ are associated with the Levi-Civita
connection related to the metric $g_{\mu\nu}$, the determinant of
which is denoted by $g$. If $\om$ is torsion-free $R = - 2 \ast \cR $.
In the presence of boundaries a York--Gibbons--Hawking-like
boundary term has to be added to the actions \eqref{eq:FOG} and
\eqref{eq:GDT}, the precise form of which depends on what kind of
variational principle one would like to employ \cite{Bergamin:2005pg}.
In the present work boundaries will not be considered and hence all
boundary terms will be dropped. Since \eqref{eq:GDT} is a standard dilaton gravity action as encountered e.g.~in low energy effective descriptions of string theory, the nomenclature ``dilaton'' for the field $X$ is evident.
In this context it should be mentioned that 
often the string dilaton field $\phi$ is employed, with
\begin{equation}
  \label{eq:dilatondilaton}
  X=e^{-2\phi}\,.
\end{equation}
This brings \eqref{eq:GDT} into the well-known form
\begin{equation}
  \label{eq:GDTotherdilaton}
  S^{(2')}=-\frac12\int_{\mathcal{M}_2} \extd^{2}x\, \sqrt{-g}\; e^{-2\phi} \left[ R + \hat{U}(\phi)\; (\nabla\phi)^{2} \; +\hat{V}(\phi) \right] \, ,
\end{equation}
where the new potentials $\hat{U}$, $\hat{V}$ are related to the old ones via
\begin{equation}
  \label{eq:newpotentials}
  \hat{U}=4 e^{-2\phi}U(e^{-2\phi})\,,\quad\hat{V}=- 2e^{2\phi}V(e^{-2\phi})\,.
\end{equation}
Two prominent examples are the Witten black hole
\cite{Witten:1991yr,Callan:1992rs} 
with
\begin{equation}
  \label{eq:potentialwittenbh}
  U(X)=-\frac{1}{X}\,,\quad V(X)=-2b^2X\,,\quad\rightarrow\quad\hat{U}(\phi)=-4\,,\quad\hat{V}=+4b^2\,,
\end{equation}
and the Jackiw-Teitelboim \cite{Teitelboim:1983ux} 
model 
with
\begin{equation}
  \label{eq:potentialsJT}
  U(X)=0\,,\quad V(X)= \frac{\La}{2} X\,,\quad\rightarrow\quad\hat{U}(\phi)=0\,,\quad\hat{V}=-\La\,.
\end{equation}
Other models are summarised in table \ref{tab:1}.

\begin{table}[t]
\centering \hspace*{-1.2truecm}
\fbox{
\begin{tabular}{|p{5.7cm}||>{$}c<{$}|>{$}c<{$}||>{$}c<{$}|} 
\hline
Model (cf.~\eqref{eq:GDT} or \eqref{eq:FOG})& U(X) & V(X) & w(X) $\,\,(cf.~\eqref{eq:w})$ \\ \hline \hline
1.~Schwarzschild  \cite{Thomi:1984na} 
& -\frac{1}{2X} & -\lambda^2 & -2\la^2\sqrt{X} \\
2.~Jackiw-Teitelboim \cite{Teitelboim:1983ux} & 0 & \frac{\Lambda}{2} X & \frac{\Lambda}{4} X^2 \\ 
3.~Witten BH/CGHS \cite{Witten:1991yr,Callan:1992rs} & -\frac{1}{X} & -2b^2 X & -2b^2X \\
4.~CT Witten BH \cite{Witten:1991yr,Callan:1992rs} & 0 & -2b^2  & -2b^2X \\
5.~Schwarzschild ${D>3}\,,$ \newline {\small ${\ \ \ \la_D^2=\frac{\la^2}{2}(D-2)(D-3)}$} & -\frac{D-3}{(D-2)X} & -\lambda_D^2 X^{(D-4)/(D-2)} & -\la_D^2\frac{D-2}{D-3} X^{(D-3)/(D-2)}\\  
6.~$(A)dS_2$ ground state \cite{Lemos:1994py} &  -\frac{a}{X} & -\frac{B}{2}X  & a\neq 2:\,\,-\frac{B}{2(2-a)} X^{2-a}\\
7.~Rindler ground state \cite{Fabbri:1996bz} & -\frac{a}{X} & -\frac{B}{2} X^a  & -\frac{B}{2} X \\
8.~BH attractor \cite{Grumiller:2003hq} & 0 & -\frac{B}{2}X^{-1} & -\frac{B}{2}\ln{X} \\ \hline
9.~All above: $ab$-family \cite{Katanaev:1997ni} & -\frac{a}{X} & -\frac{B}{2} X^{a+b} & b\neq-1:\,\,-\frac{B}{2(b+1)}X^{b+1} \\   \hline 
10.~Liouville gravity \cite{Nakayama:2004vk} & a & b e^{\al X} & a\neq-\al:\,\,\frac{b}{a+\al}e^{(a+\al)X} \\
11.~Scattering trivial boson \cite{Grumiller:2002dm} & $generic$  & UV+V'=0 & \propto X \\
12.~Scattering trivial fermion & $0$  & $const.$ & \propto X \\
13.~Reissner-Nordstr\"om \cite{Reissner:1916} & -\frac{1}{2X} & -\lambda^2 + \frac{Q^2}{X} & -2\la^2\sqrt{X}-2Q^2/\sqrt{X}\\
14.~Schwarzschild-$(A)dS$ \cite{Hawking:1982dh} & -\frac{1}{2X} & -\lambda^2 - \ell X & -2\la^2\sqrt{X} - \frac23 \ell X^{3/2} \\
15.~Katanaev-Volovich \cite{Katanaev:1986wk} & \alpha & \beta X^2 - \Lambda  & \int^X e^{\al y}(\be y^2-\La)\extd y\\
16.~Achucarro-Ortiz \cite{Achucarro:1993fd} & 0 &  \frac{Q^2}{X} - \frac{J}{4X^3} - \Lambda X &  Q^2\ln{X} + \frac{J}{8X^2} - \frac12 \La X^2 \\
17.~KK reduced CS \cite{Guralnik:2003we} & 0 & \frac12 X(c-X^2) & -\frac18 (c-X^2)^2 \\ 
18.~Symmetric kink \cite{Bergamin:2005au} & {\rm generic} & -X\Pi_{i=1}^n(X^2-X_i^2) & $cf.~\cite{Bergamin:2005au}$ \\
19.~2D type 0A/0B \cite{Douglas:2003up} & -\frac{1}{X} & -2b^2X+\frac{b^2q^2}{8\pi}  & -2b^2X+\frac{b^2q^2}{8\pi}\ln{X}\\
20.~exact string BH \cite{Dijkgraaf:1992ba} & $cf.~\cite{Grumiller:2005sq}$ & $cf.~\cite{Grumiller:2005sq}$  & $cf.~\cite{Grumiller:2005sq}$ \\
21.~KK red.~conf.~flat \cite{prep} & 0 & \frac{B}{4} \sin{(X/2)} & -\frac{B}{2} \cos{(X/2)} \\
22.~Dual model to 21 \cite{prep} & -\frac{1}{2}\tanh{(X/2)} & \frac{A}{8}\sinh{(X)} & \frac{A}{2} \cosh{(X/2)} \\
23.~Dual model to 17 \cite{prep} & -\frac{2}{X} & 2MX^3+\frac{1}{4X} & MX^2-\frac{1}{8X^2} \\
\hline
\end{tabular}
}
\caption{List of models (extending the version in \cite{Grumiller:2006rc})}
\label{tab:1}
\end{table}

Although first order gravity \eqref{eq:FOG} is not conformally
invariant, dilaton dependent conformal transformations
\begin{align}\label{eq:FOGconftrans}
X^a \mapsto \frac{X^a}{\Om}\,, && e^a \mapsto e^a \Om\,, && \om \mapsto \om + X_a e^a \frac{\extd \ln \Om}{\extd X}
\end{align}
with a conformal factor $\Om = \exp\big[\frac{1}{2}\int\limits^X (U(y)
- \tilde{U}(y)) \extd y\big]$ map a model with potentials
$(U(X),V(X))$ to one with $(\tilde{U}(X),\tilde{V}(X)= \Om^2 V(X))$.
Thus one can always transform to a conformal frame with $\tilde{U}=0$,
which considerably simplifies the classical equations of motion, do
calculations there and afterwards transform back to the original
conformal frame.  The expression
\eq{w(X) = \int\limits^X e^{Q(y)} V(y) \extd y}{eq:w}
is invariant under conformal transformations, whereas
\eq{Q(X) = \int\limits^X U(y) \extd y}{eq:wI}
captures the information about the conformal frame.

%
It turns out that there is an absolute (in space and time) conserved
quantity, 
\eq{\cC^{(g)} = e^{Q(X)}X^+X^- + w(X)\,,\quad \extd \cC^{(g)} = 0\,,}{eq:conservedquantity}
which has been found in previous second order studies of dilaton
gravity \cite{Frolov:1992xx}. 
This local gravitational mass is nothing but the Misner-Sharp mass \cite{Misner:1964je} for
spherically reduced gravity. It also exists for first order gravity
coupled to matter fields \cite{Kummer:1995qv,Kummer:1998yg}, where the
conservation law $\extd(\cC^{(g)} + \cC^{(m)}) = 0$ receives a matter
contribution.

\section{Constraint and BRST Analysis}\label{se:3}

In this section we will briefly review the constraint structure of our
model (for details see \cite{Meyer:2005fz}) and then
obtain the BRST charge needed to construct the gauge fixed and
BRST-invariant Hamiltonian, which in turn then is the starting point
for path integral quantization of \eqref{eq:start}. The constraint analysis for the
special case of massless, non self-interacting and minimally coupled
fermions was carried out in \cite{wal01}.

We will frequently denote the canonical coordinates and momenta by 
\begin{equation}
  \label{eq:newD1}
  \ol{q}^i = (\om_0,e_0^-,e_0^+)\,,\qquad q^i = (\om_1,e_1^-,e_1^+)\, \qquad p_i = (X,X^+,X^-)\,, 
\end{equation}
with $i=1,2,3$ and ($\al = 0,1,2,3$)
\begin{equation}
  \label{eq:newD2}
  Q^\al = (\chi_0,\chi_1,\chid_0,\chid_1)\,.
\end{equation}
The graded Poisson bracket is the usual one,
i.e.~$\spoiss{q^i}{p_j^\prime}=\de^i_j\de({x^1}-{x^1}^\prime)$ for the
bosonic variables and $\spoiss{Q^\al}{P_\be'} = - \de^\al_\be
\de({x^1}-{x^1}')$ for the fermionic ones. The prime denotes evaluation at
${x^1}'$, whereas quantities without prime are evaluated at ${x^1}$.

First order gravity has three gauge degrees of freedom, the local
$SO(1,1)$ symmetry and two non-linear symmetries which on-shell
correspond to space-time diffeomorphisms. Thus the system is
anticipated to possess a corresponding number of first class
constraints in the Hamiltonian formulation of the theory. Because
\eqref{eq:FOG} does not contain time (i.e.~$x^0$) derivatives of the
$\ol{q}^i$, their momenta $\ol{p}_i \approx 0$ weakly vanish.  These
are three additional first class constraints appearing in the
Hamiltonian formalism of the first order theory \eqref{eq:FOG} which
generate shifts of the $\ol{q}^i$ in the extended phase space that
includes the pair
$(\ol{q}^i,\ol{p}_i)$. 

Because the fermion kinetic term \eqref{eq:fermkin} is of first order
in the derivatives, the momenta $P_\al = \partial^L \cL / \partial
\dot{Q}^\al$ conjugate to the $Q^{\al}$ yield second class constraints
\begin{align}\nonumber
\Phi_0 &= \, P_0 + \frac{i}{\sqrt{2}}f(p_1) q^3 Q^2\approx 0,\quad & \Phi_1 &= \, P_1 - \frac{i}{\sqrt{2}}f(p_1) q^2 Q^3\approx 0,\\\label{eq:2ndclass}
\Phi_2 &= \, P_2 + \frac{i}{\sqrt{2}}f(p_1) q^3 Q^0\approx 0,\quad & \Phi_3 &= \, P_3 - \frac{i}{\sqrt{2}}f(p_1) q^2 Q^1\approx 0.
\end{align}
To reduce the phase space to the surface defined by the second class
constraints without explicitly solving them we introduce the Dirac
bracket \cite{Dirac:1996,Henneaux:1992}
\eq{
\dirac{f(x)}{g(y)} := \spoiss{f}{g} - \int \extd z \extd w \; \spoiss{f(x)}{\Phi_\al(z)} C^{\al\be}(z,w) \spoiss{\Phi_\be(w)}{g(y)}\,. }{diracbracket}
%
%
The matrix-valued
distribution 
\eq{
C^{\al\be}(x,y) = \frac{i}{\sqrt{2}f(X)}\, \left(
                      \begin{matrix}
                      0      & 0     & \frac{1}{e_1^+} & 0     \\
                      0      & 0     & 0      & -\frac{1}{e_1^-} \\
                      \frac{1}{e_1^+} & 0     & 0      & 0     \\
                      0      & -\frac{1}{e_1^-} & 0      & 0
                      \end{matrix}
                      \right) \de({x^1}-{y^1}) }{eq:Calbeinv}
is the inverse of the
Dirac matrix $C_{\al\be}(z,w) = \spoiss{\Phi_\al(z)}{\Phi_\be(w)}$, viz.
\eq{
C_{\al\be}(x,y) = i\sqrt{2}f(X) \left(
                      \begin{matrix}
                      0      & 0     & -e_1^+ & 0     \\
                      0      & 0     & 0      & e_1^- \\
                      -e_1^+ & 0     & 0      & 0     \\
                      0      & e_1^- & 0      & 0
                      \end{matrix}
                      \right) \de({x^1}-{y^1}) \,.}{eq:Calbe}

Requiring constancy of the primary first class constraints under time
evolution yields secondary first class constraints $G_i =
\dirac{\ol{p}_i}{{\cal H}'} \approx 0$ which explicitly read
\begin{align}                                                                                         \label{G1}
G_1 & =  G_1^g                                                                             \\\label{G2}
G_2 & =  G_2^g + \frac{i}{\sqrt{2}}f(X)(\chid_1 \lrpd{1} \chi_1) + e_1^+ h(X) g(\chib\chi) \\\label{G3}
G_3 & =  G_3^g - \frac{i}{\sqrt{2}}f(X)(\chid_0 \lrpd{1} \chi_0) - e_1^- h(X) g(\chib\chi)\,.
\end{align}
The constraints of the matterless theory \cite{Kummer:1996hy} are
\begin{align}\label{G1g}
G_1^g & = \partial_1 X + X^- e_1^+ - X^+ e_1^-                                            \\\label{G2g}
G_2^g & = \partial_1 X^+ + \om_1 X^+ - e_1^+ \cal V                                       \\\label{G3g}
G_3^g & = \partial_1 X^- - \om_1 X^- + e_1^- \cal V\,.
\end{align}
The Lagrangian and Hamiltonian densities of the combined system
\eqref{eq:FOG}, \eqref{eq:fermkin} and \eqref{eq:fermSI} are related
by Legendre transformation ${\cal H} = \dot{Q}^\al P_\al + p_i
\dot{q}^i - \cL$.  The Hamiltonian
%
\beqa\label{eq:constrainedhamiltonian}
H & = & \int \extd x^1 \cH = - \int \extd x^1 \ol{q}^i G_i \approx 0
\eeqa
vanishes on the constraint surface, as expected for a generally
covariant system \cite{Henneaux:1992}, and the $\ol{q}^i$ serve as
Lagrange multipliers for the secondary constraints. They form an
algebra \cite{Meyer:2005fz}
\begin{align}\label{GiGi}
\dirac{G_i}{G_i '} & =  0\,, \qquad i = 1,2,3\,, \\\label{G1G2}
\dirac{G_1}{G_2 '} & =  - G_2 \,\de({x^1}-{x^1}^\prime)\,, \\\label{G1G3}
\dirac{G_1}{G_3 '} & =  G_3 \,\de({x^1}-{x^1}^\prime)\,, \\\label{G2G3}
\dirac{G_2}{G_3 '} & =  \left[ - \sum\limits_{i=1}^3 \td{{\cal V}}{p_i} G_i + \left(g h' - \frac{h}{f} f'g'\cdot (\chib\chi)\right)G_1 \right]\de({x^1}-{x^1}^\prime)\,,
\end{align}
and are of first class because the right hand sides vanish weakly.
The $G_i$ are preserved under time evolution, $\dot{G}_i =
\dirac{G_i}{\cH'} = -{\ol{q}^j}' \dirac{G_i}{G_j'} \approx 0$, so no
ternary constraints are generated. The algebra closes with
$\de$-functions, resembling rather an ordinary gauge theory or
Ashtekar's approach to gravity
\cite{Ashtekar:1986yd} 
than the ADM approach \cite{Arnowitt:1962} whose Hamiltonian and
diffeomorphism constraints are recovered by linear combinations of the
$G_i$ and fulfil the classical Virasoro algebra
\cite{Katanaev:1994qf}. The case of minimally coupled massless
fermions \cite{wal01} without self-interaction is reproduced. If the
dilaton couplings $f,h$ are proportional to each other, only a
Thirring term $\lambda (\chib\chi)^2$ contributes to the last term in
\eqref{G2G3}. In contrast to the matterless case \cite{Grosse:1992eb}
the algebra generated by the $G_i$ and $p_i$ is no finite W-algebra
\cite{deBoer:1995nu} anymore (for a proof, cf. sec.~2.2.2 in \cite{Meyer:2006vh}).

In order to obtain a gauge fixed Hamiltonian density, we follow the
method of Batalin, Vilkovisky and Fradkin
\cite{Fradkin:1975cq} 
and first
construct the BRST charge $\Om$. With three gauge symmetries generated
by the $G_i$ the phase space has to be enlarged by three pairs of
ghosts and antighosts $(c_i, p_i^c)$ and equipped with a Poisson
structure obeying the same (anti)commutation relations for the
$q^i,p_i,Q^\al,P_\al$ as above together with $\spoiss{c^i}{{p_j^c}'} =
- \de^i_j \de({x^1}-{x^1}')$ for the ghost sector. The Dirac bracket is still
defined as in \eqref{diracbracket}, but with the new Poisson
structure. The BRST charge has to fulfil four requirements: First it
has to act on functions on the enlarged phase space through the Dirac
bracket, $\Om F(q,p,Q,P,c,p^c) := \dirac{\Om}{F}$.  Second, it has to
be nilpotent, $\Om^2 F = 0$, which by virtue of the Jacobi identity is
equivalent to
\eq{\dirac{\Om}{\Om} = 0\,.}{eq:brstnilpot}
Third, it should act on functions on the non-extended phase space as
gauge transformations, i.e.~through $G_i$ and, fourth, is required to
have ghost number one, which leads to the Ansatz $\Om = c^i G_i +
\mathrm{higher\; ghost\; terms}$. Constructed in this way it is unique
up to canonical transformations of the extended phase space
\cite{Henneaux:1992}. Evaluating \eqref{eq:brstnilpot} yields
(${C_{ij}}^k$ are the structure functions of the algebra
\eqref{GiGi}-\eqref{G2G3}, $\dirac{G_i}{G_j} = {C_{ij}}^k G_k
\de({x^1}-{x^1}^\prime)$)
\eq{ \Om = c^i G_i + \frac{1}{2}c^i c^j {C_{ij}}^k p_k^c\,.}{eq:BRSTcharge}
The homological perturbation series terminates at Yang-Mills level,
i.e.~\eqref{eq:brstnilpot} holds for \eqref{eq:BRSTcharge} without the
necessity of introducing higher order ghost terms.  BRST invariant functionals with total
ghost number zero are then sums of a BRST closed and a BRST exact part
\cite{Weinberg:1995II}. The gauge fixed Hamiltonian should be BRST
invariant and thus is of form
\eq{{\cal H}_{gf} = {\cal H}_{BRST} + \dirac{\Om}{\Psi}\,.}{eq:keinlabel}
Choosing the gauge fixing fermion \cite{Grumiller:2001ea}
\eq{\Psi = p_2^c}{eq:gaugefermion}
and ${\cal H}_{BRST} = 0$ yields the gauge fixed Hamiltonian density
\eq{{\cal H}_{gf} = \dirac{\Om}{\Psi} = -G_2 - {C_{2i}}^k c^i p_k^c}{eq:gfhamiltonian}
in Eddington-Finkelstein (or Sachs-Bondi) gauge
\eq{(\om_0,e_0^-,e_0^+) = (0,1,0)\,.}{eq:EFgauge}
The gauge fixed Lagrangian
\eq{ {\cal L}_{gf} =  \dot{Q}^\al P_\al + \dot{q}^i p_i + G_2 + p_k^c {M^k}_l c^l}{eq:gflagrangian}
contains the Faddeev-Popov operator
\eq{
M = \left(
\begin{array}{lll}
\partial_0 & 0 & \pd{\cV}{X} - \left(g h' - \frac{h}{f} f'g'\cdot (\chib\chi)\right) \\
-1 & \partial_0 & \pd{\cV}{X^+} \\
0 & 0 & \partial_0 + \pd{\cV}{X^-}
\end{array}\right)\,.
}{eq:FPoperator}

\section{Integrating out Geometry Non-perturbatively}\label{se:4}

In this section we will perform the path integration over the
(anti)ghosts $(c^i,p_i^c)$ and the geometric variables $(q^i,p_i)$
non-perturbatively. We introduce external sources for the latter and
the fermion,
\eq{\cL_{\rm src} = J^ip_i + j_i q^i + \etab\chi + \chib\eta\,.}{eq:sourcelagrangian}
The generating functional of Green functions is formally given by the
path integral with the action \eqref{eq:gflagrangian} and
\eqref{eq:sourcelagrangian} ($\cN$ is a normalisation factor)
\eq{Z[J,j,\eta,\etab] = \cN \int \cD\mu[Q,P,q,p,c,p^c] \exp{\left(i \int \extd^2x (\cL_{gf} + \cL_{\rm src}) \right)}}{eq:genfunct}
and the measure
\eq{\cD\mu[Q,P,q,p,c,p^c] = \cD\chib \cD\chi \; \prod\limits_x \frac{1}{[q^3]^2} \prod\limits_{i=1}^3 \cD p_i \cD q^i  \prod\limits_{i=0}^3 \cD P_i \delta(\Phi_i) \prod\limits_{j=1}^3 \cD c^{\,j} \cD p^c_j\,.}{eq:PImeasure}
The delta functional in the measure restricts the integration to the
surface defined by the second class constraints \cite{Henneaux:1992}.
The local measure factor for the fermion integration has been chosen
such that general covariance is retained in the quantum theory
\cite{Fujikawa:1987ie}. 
From the point of view of the phase space path integral it is for
minimal coupling coupling ($f(X) = 1$) composed of a well-known
\cite{Abers:1973qs} factor $[-g^{00}] = 2q^2/q^3$ and a factor
$\sqrt{\sdet C_{\al\be}} = (\det C_{\al\be})^{-1/2} = (4(q^2
q^3)^2)^{-1/2}$, where the latter results from rewriting the path
integral over the surface of second class constraints as a path
integral over the whole phase space
\cite{Henneaux:1992,Henneaux:1994jf}. For non-minimal coupling the
question of which measure is the ``right'' covariant one is subtle and
still not completely settled (for a review cf.~e.g.
\cite{Kummer:1999zy}).

Integrating over the ghost sector yields the functional determinant of
the Faddeev-Popov operator \eqref{eq:FPoperator},
\eq{\Delta_{\Phi\Pi} = \mathrm{Det}(\partial_0^2(\partial_0 + U(X)X^+))\,,}{eq:FPdet}
which will be cancelled during the $p_i$-integration later on.
Integration of the fermion momenta $P_i$ is trivial because of the
delta functionals in \eqref{eq:PImeasure} and the $P_i$-linearity of
the second class constraints \eqref{eq:2ndclass}, and yields an
effective Lagrangian
\eq{\cL_{\rm eff}^{(1)} = p_i \dot{q}^i + G_2 + \frac{i}{\sqrt{2}}f(p_1)\left[q^3(\chid_0\lrpd{0}\chi_0) - q^2(\chid_1\lrpd{0}\chi_1) \right] + \cL_{\rm src}\,.}{eq:Leff1}
Equation \eqref{eq:Leff1} is linear in the $q^i$, and without the
non-linearity introduced by the covariant matter measure in
\eqref{eq:PImeasure} the $q^i$-integral could be evaluated
immediately.  We thus replace \cite{Kummer:1998zs} the measure factor
$[q^3]^{-2}$ 
by introducing a new field $F$,
\begin{align}
\label{eq:treatmeasure}
Z[J^i,j_i,\eta,\etab] & = \cN \int \cD F \de\left(F - \frac{1}{i} \frac{\de}{\de j_3}\right) \tilde{Z} \\ \label{eq:ztilde}
\tilde{Z}[F,J^i,j_i,\eta,\etab] & = \int \cD \chib \cD\chi \prod\limits_x F^{-2}\, \cD p_i \cD q^i \Delta_{\Phi\Pi}\, \exp{\left(i \int \extd^2 x \cL_{\rm eff}^{(1)}\right)}\,.
\end{align}
Now we use the $q^i$-linearity of \eqref{eq:Leff1} (with \eqref{G2} and
\eqref{G2g}) which 
upon functional integration yields
three $\de$-functionals containing partial differential equations for
the $p_i$,
\begin{align}\label{eq:p1}
\partial_0 p_1 & =  j_1 + p_2  \\\label{eq:p2}
\partial_0 p_2 & =  j_2 - \frac{i}{\sqrt{2}} f(p_1) (\chid_1 \lrpd{0} \chi_1) \\\label{eq:p3}
(\partial_0 + U(p_1)p_2)p_3 & =  j_3 + \frac{i}{\sqrt{2}} f(p_1) (\chid_0 \lrpd{0} \chi_0) + h(p_1) g(\chib\chi) - V(p_1)\,.\end{align}
Performing the $p_i$-integration now amounts to solving these
equations for given currents $j_i$ and matter fields and substituting
the solutions $p_i = \hat{B}_i$ back into the effective action
obtained after $q^i$-integration, yielding an effective Lagrangian
\eq{\cL_{\rm eff}^{(2)} = J^i \hat{B}_i + \etab\chi + \chib \eta + \frac{i}{\sqrt{2}} f(\hat{B}_1) (\chid_1 \lrpd{1} \chi_1)\,.}{eq:Leff2}
During this integration the Faddeev-Popov determinant \eqref{eq:FPdet}
cancels, because the differential operators on the right hand sides of
\eqref{eq:p1}-\eqref{eq:p3} combine to a factor
$\Det(\partial_0^2(\partial_0+U(X)X^+))^{-1}$.

Equations \eqref{eq:p1} and \eqref{eq:p2} can, for general non-minimal
couplings $f(p_1)$, be solved order by order in the weak matter
approximation \cite{Meyer:2006vh}, i.e.~for matter configurations with
total energy several orders of magnitude below the Planck scale.  The
most common case of non-minimal coupling is the linear one, $f(X) =
-X$, arising from spherical reduction of in four dimensions minimally
coupled matter \cite{Balasin:2004gf}. In the weak matter approximation
it can be solved with the Ansatz
\eq{p_i = \hat{B}_i = \sum\limits_{n=0}^\infty p_i^{(n)}\ \ i=1,2\,,}{eq:piansatz}
i.e.~assuming $p_1^{(n)}$ to be of order $n$ in fermion bilinears $\chid_1 \lrpd{0} \chi_1$. The
vacuum solutions $p_i^{(0)}$ are given by \eqref{eq:B1} and
\eqref{eq:B2} below with $\ka = 0$, and the higher order terms by the
recursion relations
\begin{align}\label{eq:p2linear}
p_2^{(n)} & =  \frac{i}{\sqrt{2}}\nabla_0^{-1}\left( (\chid_1 \lrpd{0} \chi_1)p_1^{(n-1)}\right)\quad n \ge 1 \\\label{eq:p1linear}
p_1^{(n)} & =  \nabla_0^{-1} p_2^{(n)}\,.
\end{align}
With
\eq{\Qh(\hat{B}_1,\hat{B}_2) = \nabla_0^{-1} (U(\hat{B}_1)\hat{B}_2)}{eq:Qreg2}
the third equation \eqref{eq:p3} is solved by
\begin{multline}
p_3 = \hat{B}_3 = e^{-\Qh(\hat{B}_1,\hat{B}_2)}\Big[ \nabla_0^{-1} e^{\Qh(\hat{B}_1,\hat{B}_2)} \Big( j_3 - V(\hat{B}_1) + h(\hat{B}_1)g(\chib\chi) \\
+ \frac{i}{\sqrt{2}} f(\hat{B}_1) (\chid_0 \lrpd{0} \chi_0) \Big) + \tilde{p}_3 \Big]\,. \label{eq:B3}
\end{multline}
For minimal coupling $f(p_1)=-\kappa=\mathrm{const}.$ the solution
even can be given non-per\-tur\-bative\-ly,
\begin{align}\label{eq:B1}
p_1 = \Bh_1 & =  \nabla_0^{-1}(j_1 + \Bh_2) + \tilde{p}_1 \\\label{eq:B2}
p_2 = \Bh_2 & =  \nabla_0^{-1}\left( j_2 + \kappa \frac{i}{\sqrt{2}} (\chid_1 \lrpd{0} \chi_1) \right) + \tilde{p}_2\,. 
\end{align}
The quantities $\tilde{p}_i$ are homogeneous solutions of $\nabla_0
\tilde{p}_i = 0$ with the regularised time derivative $\nabla_0 =
\partial_0 -i(\mu -i \eps) =
\partial_0 -i\tilde{\mu}$, where we applied the regularisation
prescription of \cite{Kummer:1998zs}.  The integral operator
$\nabla_0^{-1}$ is the Green function of $\nabla_0$. This
regularisation provides proper infrared and asymptotic behaviour of
the Green function.  In the next section we will however use another
strategy \cite{Grumiller:2000ah} to obtain the lowest order
interaction vertices by directly imposing boundary conditions and
solving the equations \eqref{eq:p1}-\eqref{eq:p3}, such that no
additional regularisation is necessary.

Equation \eqref{eq:Leff2} as it stands is not the whole effective
action, but has to be supplemented with ambiguous terms
\cite{Kummer:1998zs,Grumiller:2000ah}. These arise from the source
terms $J^i \hat{B}_i$ in the following way: In expressions like $\int
J \nabla^{-1} A$ the inverse derivative acts after changing the order
of integration on the source $J$, giving rise to another homogeneous
contribution $\int \tilde{g} A$, while the homogeneous functions in
$A$ have been made explicit already in the solutions $\hat{B}_i$. Thus
the action gets supplemented by three terms
\begin{align}\nonumber
\cL_{\rm amb} & = \sum\limits_{i=1}^2 \tilde{g}_i K_i(\nabla_0^{-1},(\chid_1\lrpd{0}\chi_1),j_1,j_2) \\\label{eq:ambiguous}
          &   \hspace{5mm}+ \tilde{g}_3 e^{\hat{Q}} \left( j_3 - V(\hat{B}_1) + h(\hat{B}_1)g(\chib\chi) + f(\hat{B}_1) \frac{i}{\sqrt{2}} (\chid_0 \lrpd{0} \chi_0) \right)\,.
\end{align}
The expressions $K_i$ can be read off from the solutions
$\hat{B}_{1/2}$ up to the desired order in matter contributions.  The
homogeneous solutions $\tilde{g}_i$ are fixed by asymptotic conditions
on the expectation values of the Zweibeine. For instance if
$\tilde{g}_3=1$ then
\eqs{
\langle e_1^+ \rangle = \left.\frac{1}{iZ}\frac{\de}{\de j_3} e^{i\int\extd^2 x(\cL_{\rm eff}^{2} + \cL_{amb})}\right|_{j_i=J^i=0} = e^{\Qh (\Bh_1[j_i=J^i=0])}}
is just the correct asymptotic expression $e^{Q}=\sqrt{-g}$ in
Eddington-Finkelstein gauge, cf.~\eqref{eq:q3effgeom} below, if the
fermion field obeys an appropriate fall-off condition. That these
ambiguous terms are necessary and can not be omitted can also be seen
from \eqref{eq:Leff2}, which is independent of the source $j_3$.
Because of the measure factor $F^{-2}$ in \eqref{eq:treatmeasure}, the
generating functional \eqref{eq:treatmeasure} and \eqref{eq:ztilde}
would be ill-defined without the last term in \eqref{eq:ambiguous}.
Also, for the special case of the Katanaev-Volovich model without
matter the integration can be carried out in the ``natural'' order
\cite{Haider:1994cw}, i.e.  first over $p_i$ and then over $q^i$,
while never introducing sources $J^i$, yielding an effective action
exactly of the type of the $\tilde{g}_3$-term in \eqref{eq:ambiguous}.

Thus after integrating out the whole ghost and geometric sector, the
partition function reads ($\chit = \sqrt{F} \chi$)
\eq{
Z[J^i,j_i,\eta,\etab] = \cN \int \cD F \de\left(F - \frac{1}{i} \frac{\de}{\de j_3}\right) \int \cD \chitb \cD\chit \exp{\left(i \int \extd^2 x (\cL_{\rm eff}^{(2)} + \cL_{\rm amb})\right)}\,,}{eq:ztilde2}
with $\cL_{\rm eff}^{(2)}$ from \eqref{eq:Leff2} and $\cL_{\rm amb}$ from
\eqref{eq:ambiguous}.  It should be emphasised that $Z$ includes all
gravitational backreactions, because the auxiliary field $F$ upon
integration is equivalent to the quantum version of $e_1^+$.

\section{Matter Perturbation Theory}\label{se:5}

The remaining matter integration in
\eqref{eq:ztilde2} is carried out perturbatively.
One first splits the effective action \eqref{eq:Leff2} and
\eqref{eq:ambiguous} into terms independent of the fermions, in those
quadratic in the spinor components and in higher order terms
summarised in an interaction Lagrangian $\cL_{\rm int}$. The solutions
of \eqref{eq:p1}~and~\eqref{eq:p2} up to quadratic fermion terms are
for general non-minimal coupling ($B_{1/2}$ are the zeroth order
solutions \eqref{eq:B1} and \eqref{eq:B2} with $\ka = 0$)
\begin{align}\label{eq:ph1}
\Bh_1 & =  B_1 -\frac{i}{\sqrt{2}} \nabla_0^{-2} (f(B_1) (\chid_1 \lrpd{0} \chi_1)) + \cO(\chi^4) \\\label{eq:ph2}
\Bh_2 & =  B_2 -\frac{i}{\sqrt{2}} \nabla_0^{-1} (f(B_1) (\chid_1 \lrpd{0} \chi_1)) + \cO(\chi^4)\,.
\end{align}
Expanding\footnote{Prime denotes differentiation with respect to the
  argument of $U(B_1)$ and the space-time points where the functions
  are evaluated as well as the space-time integration variables are
  denoted in subscript.} \eqref{eq:Qreg2},
\begin{align}
\Qh(\hat{B}_1,\hat{B}_2) & =  Q_x(B_1,B_2) - \frac{i}{\sqrt{2}} \int_y G_{xy} f(B_{1y})(\chid_1 \lrpd{0} \chi_1)_y \label{eq:Qh} + \cO(\chi^4)\\
G_{xy} & =  \int_z \nabla_{0xz}^{-1}[U'_z B_{2z} \nabla_{0zy}^{-2} + U_z\nabla_{0zy}^{-1}]\,, \label{eq:Gxy}
\end{align}
and \eqref{eq:B3} 
yields
\begin{align}\nonumber
\Bh_{3x} & =  B_{3x} + \frac{i}{\sqrt{2}} \int_y H_{xy} f_y(B_1) (\chid_1 \lrpd{0} \chi_1)_y \\\label{eq:Bh3exp}
      &    \hspace{5mm} + e^{-Q_x} \int_y \nabla_{0xy}^{-1} e^{Q_y} \left( \frac{i}{\sqrt{2}} f(B_1) (\chid_0 \lrpd{0} \chi_0) -m h(B_1) \chib\chi\right)_y +\cO(\chi^4)\\
H_{xy} & =  e^{-Q_x}\int_z \nabla_{0xz}^{-1}e^{Q_z}\left\{ [G_{xy} - G_{zy}] (j_3 - V)_z 
       + V'_z\nabla_{0zy}^{-2}\right\} + \tilde{p}_{3x}e^{-Q_x}G_{xy}\,.\label{eq:Hxy}
\end{align}
A similar expansion of the ambiguous terms \eqref{eq:ambiguous}
yields for the whole effective action
\begin{align}\label{eq:Leffexp}
\cL_{\rm eff} & =  \cL_{\rm eff}^{(0)} + \cL_{\rm eff}^{(2)} + \cL_{\rm int} \\\label{eq:Leff0}
\cL_{\rm eff}^{(0)} & =  J^iB_i + \tilde{g}_3 e^Q(j_3-V(B_1)) + \tilde{g}_1\left( j_1 + B_2\right) + \tilde{g}_2 j_2 \\\nonumber
\cL_{\rm eff}^{(2)} & =  \frac{i}{\sqrt{2}} f(B_1) \left[(\chid_1\lrpd{1}\chi_1) - E_1^- (\chid_1\lrpd{0}\chi_1) + F^{(0)} (\chid_0 \lrpd{0} \chi_0) \right] \\ \label{eq:Leff_2}
                &    \hspace{5mm} + F^{(0)} h(B_1) m \chib\chi + \etab\chi + \chib\eta 
\end{align}
\begin{align}\nonumber
E_{1x}^-        & =  \int_y \Big[ J^1_y\nabla_{0yx}^{-2} + (J^2_y + \tilde{g}_{1y})\nabla_{0yx}^{-1} - J^3_y H_{yx} \\\label{eq:E1-}
& \hspace{1cm}
+ \tilde{g}_{3y} e^{Q_y}(G_{yx}(j_3-V)_y - V'_y\nabla_{0yx}^{-2})\Big] + \tilde{g}_{2x} \\\label{eq:E1+}
{E_{1x}^{+}}^{(0)}        &  = e^{Q_x}\left[ \int_y J^3_y e^{-Q_y} \nabla_{0yx}^{-1} + \tilde{g}_{3x} \right] =: F^{(0)}\,.
\end{align}
One recognises in $\cL_{\rm eff}^{(2)}$ the kinetic term
\eqref{eq:fermkin} of fermions on a curved background in
Eddington-Finkelstein gauge \eqref{eq:EFgauge} with a background
metric
\eq{g_{\mu\nu} = F^{(0)} \left(
\begin{array}{cc}
0 & 1 \\
1 & 2 E_1^-
\end{array}\right)_{\mu\nu} \,.
}{eq:EFmetric}
This background solely depends on sources $(j^i,J_i)$ for the
geometric variables and the zeroth order solutions $B_i$. We redefine
the interaction part of the Lagrangian density such that the
background \eqref{eq:EFmetric} depends on the full $E_1^+(x)=F(x) =
\frac{\de}{\de j_3(x)} \int \extd^2 z \cL_{\rm eff}(z)$ instead of its
matter-independent part $F^{(0)}$, i.e.~take into account
backreactions onto the metric determinant to all orders in the fermion
fields. The generating functional\footnote{There is one subtlety to
  address in connection with this interpretation of fermions on an
  effective background: The effective metric and spin connection both
  become complex if one chooses the regularisation of the inverse
  derivatives as in \cite{Kummer:1998zs}, and the kinetic term of the
  fermions \eqref{eq:fermkin} will depend on the imaginary part of the
  latter. The imaginary parts however are proportional to the
  regularisation parameter and thus vanish in the limit where the
  regularisation is removed. This subtlety plays no role for the
  method employed in subsection~\ref{subse:6-vertices}.}
\eqref{eq:ztilde} then becomes
\begin{multline}\label{eq:ztilde3}
\Zt = \exp \left( i\int \extd^2 x \cL_{\rm eff}^{(0)} + \cL_{\rm int}\left[\frac{1}{i} F^{-\frac{1}{2}} \frac{\de^L}{\de \etatb} \right] \right) \times \\
 \times \int \cD\chitb \cD \chit \exp \left(  i \int \extd^2 x  \frac{i}{2}f(B_1)  \eps_{ab} \tilde{\epsilon}^{\nu\mu} E_\mu^b (\chitb \ga^a \lrpd{\nu}  \chit) + \etatb \chit + \chitb \etat  \right) .
\end{multline}
%


\subsection{Vertices}\label{subse:6-vertices}

In this section we are interested in the gravitationally induced
scattering and thus set the sources for the geometric variables
$j_i,J^i$ to zero. For simplicity massless and non self-interacting
fermions $g(\chib\chi)=0$ are considered. The non-polynomial structure
of the effective action $\cL_{\rm
  eff}=$~\eqref{eq:Leff2}~$+$~\eqref{eq:ambiguous} gives rise to
scattering vertices with arbitrary even numbers of external legs,
which can in principle be extracted from the effective action by
expanding it order by order in spinor bilinears.  Because of the
non-locality of the effective action such calculations are very
cumbersome, so we will adopt a strategy first introduced in
\cite{Kummer:1998zs} which is based on the local quantum triviality of
first order gravity \cite{Kummer:1996hy}, i.e.~the absence of local
quantum corrections from the geometric sector of the theory. This
implies that the expectation values $\langle q^i \rangle =
\frac{\de}{\de j_i} S_{\rm eff}$ and $\langle p_i \rangle =
\frac{\de}{\de J^i} S_{\rm eff}$ fulfil the equations of motion for
the $(q^i,p_i)$ following from \eqref{eq:Leff1}
\beqa\label{eq:vertexp1}
\partial_0 p_1 & = & p_2 \\\label{eq:vertexp2}
\partial_0 p_2 & = & -2 f(p_1) \Phi_0 \\\label{eq:vertexp3}
\partial_0 p_3 & = & -U(p_1) p_2 p_3 - V(p_1) + 2 f(p_1) \Phi_1\\
\label{eq:vertexq1}
\partial_0 q^1 & = & q^3(U'(p_1)p_2p_3 + V'(p_1)) + 2f'(p_1)\big[ q^2 \Phi_0 - q^3 \Phi_1 - \Phi_2 \big] \\\label{eq:vertexq2}
\partial_0 q^2 & = & -q^1 + q^3 p_3 U(p_1)\\\label{eq:vertexq3}
\partial_0 q^3 & = & q^3p_2 U(p_1)\,,
\eeqa
with off-shell matter contributions denoted by
\eq{\Phi_0 = \frac{i}{2\sqrt{2}}(\chid_1 \lrpd{0} \chi_1)\,,\quad \Phi_1 = \frac{i}{2\sqrt{2}}(\chid_0 \lrpd{0} \chi_0)\,,\quad \Phi_2 = \frac{i}{2\sqrt{2}}(\chid_1 \lrpd{1} \chi_1)\,.}{eq:vertexmatter}
That this indeed holds can be seen from \eqref{eq:p1}-\eqref{eq:p3}
with $j_i=J^i=0$ and $g(\chib\chi)=0$, which are just
\eqref{eq:vertexp1}-\eqref{eq:vertexp3}. From
\eqref{eq:Leff2}~and~\eqref{eq:ambiguous} three four-point vertices,
\eqs{
V^{(4)} = \int_x \int_z \Big[ V_a(x,z) \Phi_0(x)\Phi_0(z)  +
V_b(x,z)\Phi_0(x) \Phi_2(z) + V_c(x,z)\Phi_0(x) \Phi_1(z)\Big]\,,
}
can be expected. Solving \eqref{eq:vertexp1}-\eqref{eq:vertexq3} with
matter contributions
\eq{\Phi_i(x) = c_i \de^2(x-y)\,,\qquad i=0,1,2}{eq:localize}
localised at a space-time point $y$ is possible. The crucial
observation now is that inserting the solutions back into the
interaction terms of \eqref{eq:Leff1}, expanding up to $\cO(c^2)$ and
reading off the corresponding coefficients is equivalent to taking
functional derivatives $\de^2 S_{\rm eff} / \de\Phi_i(x)\de\Phi_j(y)$
and thus yields the correct four-point interaction vertices.

To solve the six first order differential equations
\eqref{eq:vertexp1}-\eqref{eq:vertexq3} unambiguously
\cite{Grumiller:2002dm} one fixes the six integration constants in the
asymptotic region $x^0 > y^0$ by
imposing the following conditions:
\begin{itemize}
\item $p_1\big|_{x^0>y^0} = x^0$ and $p_2\big|_{x^0>y^0} = 1$, i.e.
  the dilaton is identified with the $x^0$-coordinate. This
  corresponds to a fixing of the residual gauge freedom
  \cite{Meyer:2006vh}. Consequently the $x^0$ direction is a ``radial''
  direction and $x^1$ corresponds to retarded time.

\item $\cC^{(g)}\big|_{x^0>y^0} = \cC_\infty$ fixes the integration
  constant in $p_3$, cf.~\eqref{eq:p3effgeom} below.

\item $q_3\big|_{x^0>y^0} = e^{Q(x^0)}$ then solves
  \eqref{eq:vertexq3} and defines the asymptotic unit of length.

\item The remaining two integration constants entering
  $q^2\big|_{x^0>y^0}$, which is the solution of the second order
  partial differential equation \eqref{eq:vertexq22}, are called
  $m_\infty$ and $a_\infty$ because for spherical reduced gravity
  (cf.~the fifth model in table~\ref{tab:1}) they correspond to the
  Schwarzschild mass and a Rindler acceleration.

\end{itemize}
The remaining three integration constants $m_\infty$, $\cC_\infty$ and
$a_\infty$ are not independent from each other, because for describing
a physical asymptotic region the solutions of
\eqref{eq:vertexp1}-\eqref{eq:vertexq3} also have to fulfil the first
order gravity constraints \eqref{G1g}-\eqref{G3g}. The Lorentz
constraint \eqref{G1g} then requires
\eq{\cC_\infty = m_\infty\,,\quad a_\infty = 0\,.}{eq:Cma}
Equation \eqref{G2g} even yields a vacuum solution for the spin
connection, which however is not necessary for finding the vertices,
and \eqref{G3g} is fulfilled identically.

Another differentiation of \eqref{eq:vertexq2} with respect to
$x^0$ and use of the other equations of motion yields a second order
differential equation for $q^2$,
\eq{\partial_0^2 q^2 = - w''(p_1) - 2 f'(p_1)\big[ q^2 \Phi_0 - q^3 \Phi_1 - \Phi_2 \big] + 2 f(p_1) q^3 U(p_1) \Phi_1\,.}{eq:vertexq22}
Solving \eqref{eq:vertexp1}-\eqref{eq:vertexp3}, \eqref{eq:vertexq3}
and \eqref{eq:vertexq22} in the vacuum regions $x^0\neq y^0$ and
patching the solutions according to their continuity properties
implied by \eqref{eq:localize} ($p_2$, $p_3$, $\partial_0 q^2$ jumping
at $x^0=y^0$ and $p_1$, $q^3$, $q^2$ continuous) yields with $h_i =
c_i \theta (y^0 - x^0) \de(x^1 - y^1)$
\begin{align}\label{eq:p1effgeom}
p_1 & =  x^0 + 2 f(y^0) (x^0-y^0)h_0 \\\label{eq:p2effgeom}
p_2 & =  1 + 2 f(y^0) h_0 \\\label{eq:p3effgeom}
p_3 & =  e^{-Q(p_1)}\Big[ m_\infty - w(p_1) + 2 f(y^0)h_0 (w(x^0) - w(y^0)) - 2 f(y^0) e^{Q(y^0)} h_1 \Big]\\\nonumber
q^2 & =  m_\infty - w(p_1) + 2 h_0 \Big[ 2 f(y^0)(w(x^0)-w(y^0)) + \big[ m_\infty f'(y^0) - (f w)'|_{y^0} \big] (x^0-y^0)\Big]\\
    &    \hspace{2,5cm}-2 (f e^Q)'|_{y^0} (x^0-y^0) h_1 -2 f'(y^0)(x^0 - y^0) h_2 \label{eq:q2effgeom}\\
q^3 & =  e^{Q(p_1)}\label{eq:q3effgeom}\,.
\end{align}
From the last two equations the asymptotic line element
\begin{equation}
  \label{eq:asymptds2}
  \extd s^2 = 2 e^{Q(x^0)} \extd x^1 \big[ \extd x^0 + (m_\infty - w(x^0)) \extd x^1 \big]
\end{equation}
reveals that $q^2|_{x^0>y^0} = m_\infty - w(x^0)$ is the Killing norm,
and its zeros correspond to Killing horizons.

\begin{figure}[t]
\centering
\includegraphics[width=0.8\textwidth]{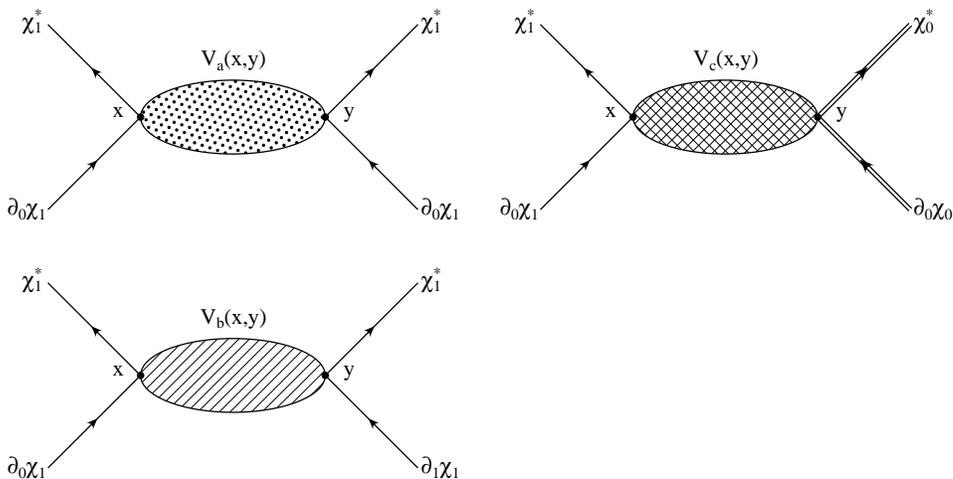}
\caption{Gravitational four-point vertices}
\label{fig:4pt}
\end{figure}

Inserting the solutions \eqref{eq:p1effgeom}-\eqref{eq:q3effgeom}
into the interaction terms of \eqref{eq:Leff1} (cf.~also
\eqref{eq:vertexmatter} and \eqref{eq:localize}),
\eq{S_{\rm int} = -2 \int_x f(p_1)\big[ q^2 \Phi_0 - q^3 \Phi_1 - \Phi_2 \big]\,,}{eq:Leff1ww}
expanding up to $\cO(c_i^2)$, replacing the coefficients $c_i$ with
the fermion bilinears
\eqref{eq:vertexmatter} 
and integrating over $y$ yields three four-point vertices (depicted in
figure~\ref{fig:4pt}), namely the symmetric one
\begin{multline}
V_a = -4 \int\limits_x \int\limits_y \Phi_0(x) \Phi_0(y) \theta(y^0-x^0) \de(x^1 - y^1) f(x^0) f(y^0) \times \\
\times \Big[ 2(w(x^0)-w(y^0)) - (x^0-y^0)(w'(x^0) + w'(y^0)) \\
- (x^0 - y^0) \Big(\frac{f'(x^0)}{f(x^0})(w(x^0) - m_\infty) + \frac{f'(y^0)}{f(y^0)}(w(y^0) - m_\infty)\Big)\Big]
\label{eq:Va}
\end{multline}
and two non-symmetric ones
\beqa
V_b & = & -4\int\limits_x \int\limits_y \Phi_0(x) \Phi_2(y) \de(x^1 - y^1) |x^0-y^0| f(x^0) f'(y^0)\label{eq:Vc}\\
V_c & = & -4\int\limits_x \int\limits_y \Phi_0(x) \Phi_1(y) \de(x^1 - y^1) |x^0-y^0| f(x^0) (f e^Q)'|_{y^0}\,.\label{eq:Vb}
\eeqa
Interestingly, the vertices $V_a$ and $V_b$ are the same as for a real
scalar field coupled to first order gravity \cite{Grumiller:2002dm}, and only $V_c$ is new. They
share some properties with the scalar case, namely
\begin{enumerate}
\item They are local in the coordinate $x^1$, and non-local in $x^0$.
\item They vanish in the local limit ($x^0\rightarrow y^0$) and $V_b$ vanishes for minimal coupling.
\item They respect a $\mathbb{Z}_2$ symmetry $f(X)\mapsto - f(X)$.
\item The symmetric vertex depends only on the conformally invariant
  combination $w(X)$ (cf.~\eqref{eq:w}) and the asymptotic value
  $m_\infty$ of the geometric part of the conserved quantity
  \eqref{eq:conservedquantity}. $V_b$ is independent of $U$, $V$ and
  $m_\infty$. Thus if $m_\infty$ is fixed in all conformal frames both
  vertices are conformally invariant.
\end{enumerate}
By contrast, the new vertex $V_c$ is not conformally invariant.

\subsection{Asymptotic Matter States}\label{subse:6-states}

For calculation of the S-matrix one needs to determine the quantities
$\Phi_i$ in \eqref{eq:Va}-\eqref{eq:Vc} from the asymptotic matter
states. They fulfil the Dirac equation
\eq{\Dir \chi = iE_a^\mu \ga^a \nabla_\mu\chi = 0\,, \quad \nabla_\mu = \partial_\mu - \frac{1}{2} \omh_\mu \ga_\ast}{eq:diracequation}
on the asymptotic background \eqref{eq:asymptds2}.  If they form a
complete and (in an appropriate sense) normalisable set, an asymptotic
Fock space can be constructed. The components of the ``Dirac
equation'' obtained from \eqref{eq:fermkin} decouple,
\beqa\label{eq:diracchi0}
\partial_0 \chi_0 & = & -\frac{1}{2}\left(\frac{f'(x^0)}{f(x^0)} + U(x^0)\right) \chi_0 \\\label{eq:diracchi1}
(\partial_1 - q^2(x^0) \partial_0) \chi_1 & = & \frac{1}{2}\left(\frac{f'(x^0)}{f(x^0)} q^2(x^0) +  {q^2}'(x^0) \right)\chi_1\,.
\eeqa
These equations can be written in the standard form
\eqref{eq:diracequation}, with a spin connection
\beqa \label{eq:om0f}
\omh_0 & = & -U(x^0) - \frac{f'(x^0)}{f(x^0)} \\\label{eq:om1f}
\omh_1 & = & -U(x^0) q^2(x^0) - {q^2}'(x^0) -2 \frac{f'(x^0)}{f(x^0)}q^2(x^0)\,.
\eeqa
Non-minimal coupling thus enters the connection by a torsion-like
contribution. For minimal coupling $f'(X)=0$ \eqref{eq:om0f} and
\eqref{eq:om1f} reduce to the Levi-Civita connection $\tilde{\om} =
e_a \ast(\extd e^a)$ calculated from \eqref{eq:asymptds2}, which
reflects the fact that in two dimensions the fermions do not couple to
the spin connection and thus do not feel torsion.
Iterating \eqref{eq:diracequation} 
yields
%
\eq{\Dir^2 \chi = - g^{\mu\nu} \nabla_\mu \nabla_\nu\chi - \frac{R}{4}\chi + e_a^\mu (\ast T^a) \nabla_\mu \chi\,,}{eq:diraciterate}
with the torsion two-form $T^a = \extd e^a + \omh \eps^a{}_b \wedge e^b$.

The equation for $\chi_1$ is conformally invariant, 
but the one for $\chi_0$ contains the potential $U(X)$ and thus
transforms while changing between conformal frames. To solve
\eqref{eq:diracchi1} one introduces coordinates ($C_\pm$ are
constants)
\beqa v_\pm(x^0,x^1) & = & x^1 \pm \int\limits^{x^0} \frac{\extd
  z}{q^2(z)} + C_\pm
\label{eq:vpmcoordinate} \eeqa
with lower integration limits such that the integrals are defined,
which is the case in the asymptotic region \eqref{eq:asymptds2}
outside horizons. In this way both coordinates are locally orthogonal
to each other, $\partial_{v_\pm} v_\mp = 0$. With the Ansatz $\chi_i =
\cR_i e^{i\phi_i}$ one finds that $\cR_0$ has to fulfil
\eqref{eq:diracchi0}, $\cR_1$ obeys
\eq{\partial_{v_-} \cR_1 = \frac{1}{4}\left(\frac{f'(x^0)}{f(x^0)} q^2(x^0) + {q^2}'(x^0) \right)\cR_1\Big|_{x^0=x^0(v_- ,x^1)}\,,}{eq:cR1v}
and the phases $\phi_i$ obey $\partial_0 \phi_0 = 0$ and
$\partial_{v_-}\phi_1 = 0$. Solutions can be given explicitly even for
general non-minimal coupling,
\beqa\label{eq:solChi0}
\chi_0(x^0,x^1) & = & \frac{F(x^1)}{\sqrt{f(x^0)}}\, \exp \Big[{i\phi_0(x^1)-\frac{Q(x^0)}{2}}\Big]\\
\chi_1(x^0,x^1) & = & \tilde{F}(v_+)e^{i\phi_1(v_+)} \times\nonumber\\ 
                &   & \hspace{-2cm}\times \exp\left[ \frac{1}{4} \int\limits^{v_-} \extd v_-' \left( \frac{f'(x^0)}{f(x^0)}q^2(x^0)  + {q^2}'(x^0) \right)\Big|_{x^0 = x^0(v_-',x^1)}\right]\,,\label{eq:solChi1}
\eeqa
with $F(x^1)$ and $\tilde{F}(v_+)$ being arbitrary real integration
functions.  The conformal transformation properties of these solutions
are consistent with table~\ref{tab:conf} below: \eqref{eq:solChi0}
includes a factor $e^{-Q/2}$ and thus transforms with weight $-1$,
while \eqref{eq:solChi1} only depends on the conformally invariant
combination $e^{Q} V$ and the mass $m_\infty$ of the asymptotic
space-time.  Thus if $m_\infty$ is fixed for all conformal frames, the
solution for $\chi_1$ is conformally invariant. The fermion bilinears
\eqref{eq:vertexmatter} which enter the tree-level scattering vertices
\eqref{eq:Va}-\eqref{eq:Vc} read on-shell
\eq{\Phi_0(x) = - \frac{|\chi_1|²}{\sqrt{2}}\frac{\phi_1'(v_+)}{q^2(x^0)} = \frac{\Phi_2(x)}{q^2(x^0)}\,,\qquad \Phi_1(x) = 0\,. }{eq:vertexmatteronshell}

\subsection{One-Loop Effects}\label{subse:6-1l}

In the following we consider massless and minimally coupled ($f(X) =
h(X) \equiv 1$) fermions. A self interaction term $g(\chib\chi) =
\lambda(\chib\chi)^2$ can be rewritten by introducing an auxiliary vector
potential in the action
\eq{S_{SI}[\chi,g,A] = \frac{\lambda}{2} \int\extd^2 x \left( F A_a^2 + 2
    A_a \chitb \ga^a \chit \right)\,,}{eq:rewrThirring}
which is integrated over in the path integral. The last term is
absorbed into the Dirac operator, such that the fermion couples in the
standard minimal way to the vector potential. After replacing
$\partial_\mu$ in \eqref{eq:ztilde3} with the metric compatible and
torsion-free covariant derivative $\nabla_\mu$ partially integrating
the kinetic term in \eqref{eq:ztilde3}, completing the square and
evaluating the Gaussian integral yields
\begin{multline}
\Zt = \exp{\left(i\int \extd^2 x \left(\cL_{\rm eff}^{(0)} + \cL_{\rm int}\left[-i F^{-\frac{1}{2}} \frac{\de^L}{\de \etatb} \right]\right)\right)} \\
\int \cD A \, \Det \Dir \, \exp{\left(i\int \extd^2 x \left( \frac{\lambda}{2} F A_a^2 - \etatb \Dir^{-1} \etat\right)\right)}\,, \label{eq:ztilde4}
\end{multline}
with the Dirac operator (cf.~\eqref{eq:diracequation}) $\Dir = i
E_a^\mu \ga^a (\nabla_\mu -i \lambda A_\mu)\,.$
The determinant of the Dirac operator is most easily calculated in
Euclidean space (where $\Dir$ is essentially self-adjoint) using
zeta-function regularisation and heat kernel methods
\cite{Vassilevich:2003xt}. The square of the Euclidean Dirac operator
(for Euclidean $\ga$-matrix conventions see \ref{se:2}) is of Laplace
type,
\eq{\Dir^2 = -(g^{\mu\nu} \nabla_\mu\nabla_\nu + E)\,,\quad E = \frac{R}{4} + \frac{\lambda}{2} \ga_\ast
\epsilon^{\mu\nu}\cF_{\mu\nu}\,.}{eq:diracsquare}
Here $\cF = \extd A$ denotes the electromagnetic field strength and
$R$ is the curvature scalar with respect to the Levi-Civita
connection. The conformal anomaly has the same value as in the scalar
case\footnote{The field strength term in $E$ does not contribute
  because of ${\rm tr_{\complexc^2}}(\ga_\ast) = 0$.},
\eq{T_\mu^\mu := g^{\mu\nu} \langle T_{\mu\nu}\rangle\, = -\frac{R}{24\pi}\,.}{eq:confanomaly}
The chiral symmetry also becomes anomalous at one-loop level, yielding
the consistent chiral anomaly
\eq{\cA(\varphi) = \de_\varphi W^{\rm ren} = - \frac{\lambda}{2\pi}\int \extd^2 x \sqrt{g} \varphi \epsilon^{\mu\nu} \cF_{\mu\nu}\,,}{eq:chiralanomaly}
where $\de_\varphi$ is an infinitesimal chiral transformation and
$W^{\rm ren}$ denotes the zeta-function renormalised one-loop
effective action. Note that although the mass-like coupling of the
vector potential to the auxiliary field $F$ in \eqref{eq:ztilde4}
breaks U(1) gauge invariance, the determinant of the Dirac operator is
still gauge invariant and therefore the methods above are applicable.

Both anomalies can be integrated (with initial condition $W[g_{\mu\nu}
= \eta_{\mu\nu}] = 0$ for the conformal anomaly), yielding a one-loop
effective action comprising a Polyakov \cite{Polyakov:1981rd} and a
Wess-Zumino part \cite{Wess:1971yu}
\beqa
W_{\rm 1loop} & = & -\ln \Det \Dir = W_{\rm Pol} + W_{\rm WZ} \label{eq:1loopaction}\\
W_{\rm Pol}   & = & \frac{1}{96\pi} \int\limits_{\cM} \extd^2 x \sqrt{-g} R \frac{1}{\Delta} R \label{eq:Polyakovaction}\\
W_{\rm WZ}    & = & \frac{\lambda}{4\pi} \int \extd^2 x \sqrt{-g} (\ast \cF)\frac{1}{\Delta}(\ast \cF) \label{eq:WessZumino}\,,
\eeqa
with the Laplacian on 0-forms $\Delta = \frac{1}{\sqrt{-g}}
\partial_\mu \sqrt{-g}g^{\mu\nu}\partial_\nu$ and the Green function
of the Laplacian defined as $\Delta_x \Delta^{-1}_{xy} = \de^{(2)}(x-y)$.

What remains to be evaluated is the path integral over the vector
potential in \eqref{eq:ztilde4}, now with the highly non-local
integrand \eqref{eq:WessZumino}. The nature of the application should
thus decide whether it is favourable to use this form, or to treat the
Thirring term directly as an interaction vertex. On the other hand it
may well be that the Wess-Zumino action becomes local in some special
gauges, as happens for the Polyakov action in conformal gauge.
Otherwise the Polyakov and Wess-Zumino action can be written in local
form 
\beqa\label{eq:WPollocal}
W_{\rm Pol} & = & \frac{1}{48\pi}\int \extd^2 x \sqrt{-g} \left[ \frac{1}{2} (\nabla \Phi)^2 + \Phi R\right] \\\label{eq:WZlocal}
W_{\rm WZ}  & = & \frac{\lambda}{2\pi}\int \extd^2 x \sqrt{-g} \left[ \frac{1}{2} (\nabla Y)^2 + Y (\ast \cF)\right]
\eeqa
by introducing two auxiliary scalars $\Phi$ and $Y$ which have to be
integrated over in the path integral ~\eqref{eq:ztilde4}.  These
expressions coincide with the ones in \cite{Nojiri:1992st,Ori:2001xc}
up to notational differences.

\section{Discussion \& Conclusions}\label{se:6}

One of the main results of this work is the non-perturbative and
background independent quantization (in the sense that no split of the
metric into a fiducial and a fluctuation part is assumed) of the
dilaton gravity sector of the theory. Although fixing the residual
gauge freedom by imposing asymptotic boundary conditions on the
momenta $p_i$ as in section~\ref{subse:6-vertices}
(cf.~\eqref{eq:p1effgeom}-\eqref{eq:p3effgeom}) determines the
geometry in the asymptotic region, the metric is still subject to
quantum fluctuations in the interior of the space-time manifold.  This
results in an effective background consistently including quantum
backreactions onto the geometry. Another key result is the
perturbative quantization of the Dirac field in that framework. The
gravitationally induced four-point vertices as well as asymptotic
fermion states have been calculated, both of which are necessary
prerequisites for S-matrix calculations.  Finally, known results for
the one-loop effective action have been recovered.

\subsection{Conformal Properties of the Effective Action}

\newcommand{\weight}{\sigma}

The action of conformal transformations $g_{\mu\nu} \mapsto \Om^2
g_{\mu\nu}$ after fixing the gauge and specifying asymptotic values of
the $p_i$ is slightly non-trivial and requires some discussion even
for massless non self-interacting fermions. This is necessary to
understand the conformal properties of the vertices and the scattering
matrix. By assumption it will be required that the asymptotic values
$\tilde{p}_i$ and the gauge fixing conditions \eqref{eq:EFgauge} are
invariant. This implies in particular that neither $e_0^-$ nor $X^+$
transform and that $e_1^+$ has the same conformal weight as the
metric. Furthermore, the conserved quantity
\eqref{eq:conservedquantity} is required to be conformally invariant.
This leads to different conformal weights (listed in
table~\ref{tab:conf}) as compared to the situation before gauge
fixing, equation \eqref{eq:FOGconftrans}. The conformal weight
$\weight(A)$ of a field monomial $A$ transforming homogeneously is
defined by its transformation behaviour $A \mapsto \Om^{\weight(A)}
A$. Conformal weights thus add for products of field monomials,
$\weight(AB) = \weight(A) + \weight(B)$.
\begin{table}[h]
\centering
\begin{tabular}{|c|c|p{4cm}|c|c|}\hline
Weight 2 & Weight 1 & Weight 0 & Weight -1 & Weight -2 \\ \hline
$g_{\mu\nu}$, $e_1^+$, $J^3$ & $\eta_1$ & $e_0^\pm$, $e_1^-$, $X$, $X^+$, $\tilde{p}_i$, $\tilde{g}_i$, $\chi_1$, $J^{1/2}$, $j_{1/2}$, $\eta_0$, $A^+$ & $\chi_0$ & $X^-$, $j_3$, $A^-$ \\ \hline
\end{tabular}
\caption{Conformal weights for Eddington-Finkelstein gauge}
\label{tab:conf}
\end{table}

\noindent The gauge fixed spin connection transforms inhomogeneously,
\begin{equation}
  \label{eq:fer1}
  \om_0\to\tilde{\om}_0=\om_0=0\,,\quad \om_1\to\tilde{\om}_1=\om_1+(X^+e_1^- + X^-e_1^+)\frac{\extd\,\ln{\Om}}{\extd X}\,,
\end{equation}
and the dilaton potentials transform as described in section
\ref{se:2} below equation~\eqref{eq:FOGconftrans}. Consequently the
effective action \eqref{eq:Leff2} and \eqref{eq:ambiguous} is
conformally invariant at tree-level, i.e.~before performing the path
integration over $\chi$ and thus before taking into account quantum
corrections from the fermions.  For \eqref{eq:Leff2} and the first two
ambiguous terms \eqref{eq:ambiguous} this is evident as all terms are
invariant by themselves. The third term in \eqref{eq:ambiguous}
requires some explanation: The four contributions in the bracket have
to transform with a weight opposite to $e^Q$; obviously, this is true
for the first two entries; for the third it holds if and only if the
mass term is absent; for the last term it holds because $\chi_0$ has
an appropriate weight and the derivative terms cancel.

\begin{figure}[t]
\centering
\includegraphics[width=0.8\linewidth]{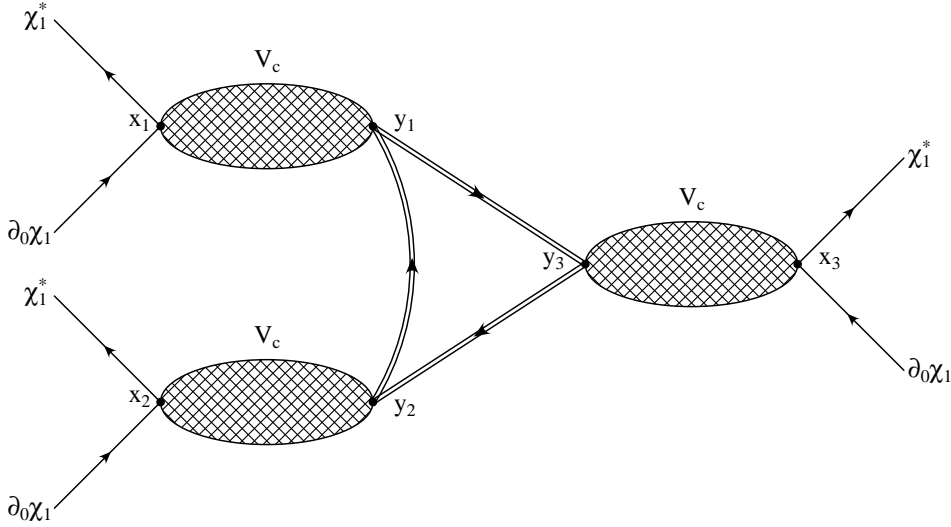}
\caption{Possibly conformally non-invariant one-loop diagram}
\label{fig:fermion6pt}
\end{figure}

For four-fermi scattering the tree-level S-matrix is indeed
conformally invariant: Although the new vertex \eqref{eq:Vb}, by
contrast to the two known ones \eqref{eq:Va} and \eqref{eq:Vc}, is not
invariant, the vanishing of the fermion bilinear $\Phi_1$
(cf.~\eqref{eq:vertexmatteronshell}) implies that the four-fermi
Feynman diagram corresponding to $V_c$ vanishes after attaching
external legs, as well as all (tree-level and loop) diagrams with two
outer $\chi_0$-legs attached to the same vertex $V_c$. This argument
extends to all scattering vertices with external $\chi_0$-legs as all
of them are generated by the last term in \eqref{eq:ambiguous} and
thus depend on $\Phi_1$. Diagrams like figure~\ref{fig:fermion6pt}
with internal $\chi_0$-propagators could contribute to the S-matrix in
a conformally non-invariant way. However, they are always (at least)
one-loop diagrams and at one-loop level conformal invariance is broken
already by the conformal anomaly \eqref{eq:confanomaly}, such that for
one-loop effects a specific choice of the potential $U$ is necessary.
In particular, two conformally related models \eqref{eq:GDT} with the
same $w$ (cf.~\eqref{eq:w}) but different $U$ lead to different
results for certain physical observables, for instance the specific
heat of black holes \cite{Grumiller:2003mc}.

The occurrence of these new interactions also imposes a stronger
requirement on scattering triviality, namely minimal coupling and
$U(X)=0$ and $V(X) = {\rm const}$, in contrast to the scalar case
where $w'(X)={\rm const.}$ was sufficient for the vanishing of all
tree-level vertices \cite{Grumiller:2002dm}. The CGHS and Rindler
ground state models (cf.  table~\ref{tab:1}) thus could exhibit
non-trivial fermion scattering.

\subsection{Comments on Bosonisation}

Frolov, Kristjansson and Thorlacius exploited bosonisation of
minimally coupled fermions to study the semi-classical geometry of
charged Witten black holes \cite{Frolov:2005ps,Frolov:2006is,Thorlacius:2006tf}. This
was possible because they quantized the fermions on a fixed background
and assumed backreactions to be small. In the present work we have
avoided such a split and integrated out geometry exactly. We collect
now some evidence in favour of bosonisation in this more general
context, but we emphasise that it is by no means conclusive.

The constraint algebra \eqref{GiGi}-\eqref{G2G3} resembles the one for
non-minimally coupled scalar matter \cite{Grumiller:2002nm}. In fact,
if the coupling functions in \eqref{eq:fermkin} and \eqref{eq:fermSI}
coincide, $f(X)=h(X)$, and only the Thirring term
$g=\lambda(\bar{\chi}\chi)^2$ is present in \eqref{eq:fermSI}, then
the structure functions of the two constraint algebras are {\em
identical}, provided that
\begin{equation}
  \label{eq:boson1}
  \phi^\pm = j^\pm \qquad\Rightarrow\qquad \epsilon^{\mu\nu}(\partial_\mu\phi)e_\nu^\pm = \bar{\chi}\ga^\pm\chi
\end{equation}
where $\phi^\pm=\ast(\extd\phi\wedge e^\pm)$ are the (anti-)self dual
components of the scalar field $\phi$ and
$j^\pm=\bar{\chi}\ga^\pm\chi$ are the (anti-)chiral fermion
currents.\footnote{The coupling constant $\lambda$ is absorbed in the
  corresponding coupling function $\tilde{f}(X)\propto f(X)$
  multiplying the kinetic term of the scalar field.} It should be
noted that \eqref{eq:boson1} is the curved version of Coleman's
bosonisation map $j_\mu \propto -\epsilon_\mu{}^\nu
\partial_\nu \varphi$ \cite{Coleman:1975pw}.  

The generating
functional \eqref{eq:ztilde4} also exploits the identification
\eqref{eq:boson1} since \eqref{eq:rewrThirring} yields $\chib \ga^a
\chi = j^a = - A^a$ for the bosonic auxiliary field $A_a$.
This trick is applicable even after quantizing the dilaton gravity sector
because the effective action \eqref{eq:Leff2} and \eqref{eq:ambiguous}
still allows the interpretation of fermions propagating on an
effective background including quantum backreactions on the geometry.

Using perturbation theory in the matter sector we found that two of
the lowest order vertices, \eqref{eq:Va} and \eqref{eq:Vc}, coincide
with the vertices found in the scalar theory \cite{Grumiller:2002dm},
and the new vertex \eqref{eq:Vb} does not contribute to tree-level
Feynman
diagrams. 

In conclusion, the results of this work indicate the presence of
bosonisation in 2D quantum dilaton gravity beyond fixed background
quantization.  For further investigation of bosonisation along the
lines of the present work it is necessary to calculate S-matrix
elements and compare with the scalar case.

\subsection{Outlook}

As all tools are available now, the next natural step is the
evaluation of the four-fermi S-matrix elements, by analogy to
minimally \cite{Grumiller:2000ah} and non-minimally
\cite{Fischer:2001vz} coupled scalar fields. Similar effects that
led to the prediction of decaying s-waves ought to be present for
fermions. Moreover, by analogy to \cite{Grumiller:2001rg} one may also
check CPT invariance.  Note that although spherical reduction of
four-dimensional Einstein-Hilbert gravity with a minimally coupled
Dirac fermion \cite{Balasin:2004gf} yields spherical reduced gravity
with two non-minimally coupled Dirac fermions plus local interaction
terms, the quantization procedure given in this work is still
applicable.
The additional Dirac spinor and the corresponding interactions will,
however, lead to additional matter terms in \eqref{eq:vertexp1}-\eqref{eq:vertexq3} and thus
give rise to new gravitational interaction vertices involving both
fermion generations. From the four-dimensional point of view only such a setting is truly spherically symmetric. Obviously, the calculation of scattering amplitudes will be more involved and new physcial effects may be expected due to the presence of these additional interactions.

We mention again that conformal invariance of S-matrix elements, which
holds at tree-level to all orders, is no longer true for one-loop
amplitudes. By contrast to the case of scalar matter there exists a
tree level vertex breaking conformal invariance, $V_c$ in our
nomenclature (cf.~\eqref{eq:Vb}), which however cannot contribute to tree-level amplitudes
as a consequence of conformal invariance of the effective action
\eqref{eq:Leff2} and \eqref{eq:ambiguous}.

\begin{figure}[t]
\centering
\includegraphics[width=0.35\linewidth]{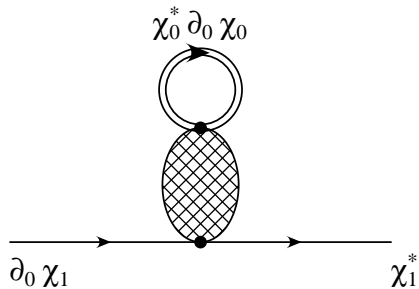}
\caption{One-loop correction from the new vertex $V_c$}
\label{fig:3}
\end{figure}

At one-loop level corrections to the specific heat of black hole
solutions are expected. In particular, the CGHS model
\cite{Callan:1992rs} -- which exhibits scattering triviality for
scalar fields but not for fermions -- should allow a straightforward
application of our results, by analogy to \cite{Grumiller:2003mc}. In
addition to the contributions from the Polyakov loop calculated there
corrections from the Feynman diagram depicted in figure \ref{fig:3}
will arise. If the Thirring term is present the one-loop effective
action receives an additional contribution \eqref{eq:WessZumino} from
the chiral anomaly.

Higher order vertices can
be obtained with essentially the same algorithm as above, and it could
be of interest to implement such algorithms in some computer algebra
system, in order to obtain expressions for arbitrary $2n$-point
vertices. With such a result available possible (partial) resummations
of the perturbation series could allow a discussion of bound states,
i.e., black holes as long-living intermediate states, in a
Bethe-Salpeter like manner. Other possible applications are outlined
in \cite{Grumiller:2002nm,Grumiller:2004yq}.


Finally, we stress that the exact path integration over geometry performed
in this work appears to be possible in two dimensions only. Despite of this fact we
can point out some observations which may serve as lessons for
higher-dimensional quantum gravity:
\begin{enumerate}

\item It has been crucial from a technical point of view to employ the
Vielbein formalism rather than the metric formalism.

\item Choosing Minkowskian signature from the beginning has led to
simplifications by enabling us to choose light-cone gauge.

\item Our path integral contains a sum over all configurations, including
singular ones, and also a sum over both ``East Coast'' and ``West Coast''
sign conventions.
The sum over singular configurations includes those where $e_1^+$
 vanishes, which up to a sign is the determinant of the Zweibein in our gauge. The sum
over both signatures is a consequence of summing over
configurations with either positive or negative $e_1^+$ and, for spherically
reduced gravity, additionally summing over positive and negative values
of the dilaton field.
So in the terminology of \cite{Duff:2006iy} our
quantization procedure may be dubbed ``bicoastal''.

\item The asymptotic conditions we had to impose in order to fix the residual
gauge freedom led to a breaking of ``bicoastalness'' and to a unique
asymptotic line element. However, the concept of some smooth ``background
geometry'' makes sense only in the asymptotic region.

\item From experience with the case of real scalars coupled to dilaton
			gravity and from the fact that we only use standard quantum field theory
			methods we expect the S-matrix to be unitary despite the emergence of
			``virtual black hole'' intermediate states \cite{Grumiller:2004yq,Meyer:2006vh}.

\end{enumerate}

\section*{Acknowledgements}

We are deeply indebted to Dima Vassilevich for numerous helpful
discussions. Additionally we are grateful to Luzi Bergamin, Max
Kreuzer and Wolfgang Kummer for helpful remarks.

This work has been supported by project J2330-N08 of the Austrian
Science Foundation (FWF), by project GR-3157/1-1 of the German
Research Foundation (DFG) and during its final stage by project MC-OIF 021421 of the European Commission. 
RM has been supported financially by the
MPI MIS Leipzig and expresses his gratitude to J.~Jost in this regard.  We
would like to thank the Vienna University of Technology for the
hospitality while part of this work was conceived.

\section*{References}


\providecommand{\href}[2]{#2}\begingroup\raggedright\endgroup

\end{document}

%% file: macros.tex
\newcommand{\beqs}{\begin{equation*}}
\def\beq{\begin{equation}}

\newcommand{\eeqs}{\end{equation*}}
\def\eeq{\end{equation}}

\newcommand{\beqas}{\begin{eqnarray*}}
\newcommand{\beqa}{\begin{eqnarray}}

\newcommand{\eeqas}{\end{eqnarray*}}
\newcommand{\eeqa}{\end{eqnarray}}

\newcommand{\eqs}[1]{\begin{equation} #1 \end{equation}}

\newcommand{\eq}[2]{\begin{equation} #1 \label{#2} \end{equation}}

\newcommand{\eps}{\varepsilon}
\newcommand{\al}{\alpha}
\newcommand{\be}{\beta}
\newcommand{\ga}{\gamma}
\newcommand{\de}{\delta}
\newcommand{\om}{\omega}
\newcommand{\ka}{\kappa}
\newcommand{\la}{\lambda}

\newcommand{\Om}{\Omega}

\newcommand{\La}{\Lambda}

\newcommand{\blist}{\begin{itemize}}

\newcommand{\elist}{\end{itemize}}

\providecommand{\href}[2]{#2}

\DeclareFontFamily{OT1}{rsfs}{}
\DeclareFontShape{OT1}{rsfs}{m}{n}{ <-7> rsfs5 <7-10> rsfs7 <10->rsfs10}{} 
\DeclareMathAlphabet{\mycal}{OT1}{rsfs}{m}{n}

\newcommand{\lrpd}[1]{\overleftrightarrow{\partial_{#1}}}

\newcommand{\chib}{\overline{\chi}}
\newcommand{\chid}{\chi^*}
\newcommand{\chitb}{\overline{\tilde{\chi}}}
\newcommand{\chit}{\tilde{\chi}}
\newcommand{\etatb}{\overline{\tilde{\eta}}}
\newcommand{\etat}{\tilde{\eta}}
\newcommand{\complexc}{{\mathbb C}}
\def\etab{\overline{\eta}}

\usepackage{slashed}
 
\def\Det{{\rm Det}}
\def\Dir{\slashed{D}}

\def\Zt{\tilde{Z}}
\def\Qh{\hat{Q}}
\def\Bh{\hat{B}}

\def\ol{\overline}
\def\cV{{\cal V}}
\def\cL{{\cal L}}
\def\cH{{\cal H}}
\def\cD{{\cal D}}
\def\cN{{\cal N}}
\def\cM{{\cal M}}

\def\cR{{\cal R}}
\def\cC{{\cal C}}
\def\cO{{\cal O}}
\def\cA{{\cal A}}
\def\cF{{\cal F}}

\def\extd{{\rm d}}

\newcommand{\pd}[2]{\frac{\partial {#1}}{\partial {#2}}}

\newcommand{\td}[2]{\frac{\mathrm{d} #1}{\mathrm{d} #2}}

\newcommand{\spoiss}[2]{ \{ #1 , #2 \}}
\newcommand{\dirac}[2]{\{ #1 , #2 \}^\ast}

\def\sdet{{\rm sdet}}

\def\atbdry{\Big|_{\partial \cM}}
\def\atbdry0{\Big|_{\partial \cM_0}}
\def\atbdry1{\Big|_{\partial \cM_1}}

\def\omh{\hat{\om}}